\documentclass[aps,prx,superscriptaddress,reprint]{revtex4-2}
\usepackage{graphicx}
\usepackage{svg}
\usepackage{hyperref}
\usepackage{amsmath,amssymb}
\usepackage{color}
\usepackage{newtxtext,newtxmath}
\definecolor{myblue}{rgb}{0,0,1}
\begin{document}

\title{Beyond overcomplication: a linear model suffices to decode hidden structure-property relationships in glasses}

\author{Chenyan Wang}
\thanks{These authors contributed equally to this work.}
\affiliation{School of Physics, Peking University, Beijing 100871, People's Republic of China}

\author{Mouyang Cheng}
\thanks{These authors contributed equally to this work.}
\affiliation{School of Physics, Peking University, Beijing 100871, People's Republic of China}

\author{Ji Chen}
\email{ji.chen@pku.edu.cn}
\affiliation{School of Physics, Peking University, Beijing 100871, People's Republic of China}
\affiliation{Interdisciplinary Institute of Light-Element Quantum Materials and Research Center for Light-Element Advanced Materials, Peking University, Beijing 100871, People's Republic of China}
\affiliation{State Key Laboratory of Artificial Microstructure and Mesoscopic Physics, Frontiers Science Center for Nano-Optoelectronics, Peking University, Beijing 100871, People's Republic of China}

\begin{abstract}
Establishing reliable and interpretable structure-property relationships in glasses is a longstanding challenge in condensed matter physics. 
While modern data-driven machine learning techniques have proven highly effective in establishing structure-property correlations, many models are criticized for lacking physical interpretability and being task-specific.  
In this work, we identify an approximate linear relation between structure profiles and disorder-induced responses of glass properties based on first order perturbation theory.
We analytically demonstrate that this relationship holds universally across glassy systems with varying dimensions and distinct interaction types.
This robust theoretical relationship motivates the adoption of linear machine learning models, which we show numerically to achieve surprisingly high predictive accuracy for structure-property mapping in a wide variety of glassy materials.
We further devise regularization analysis to further enhance the interpretability of our model, bridging the gap between predictive performance and physical insight. 
Overall, this linear relation establishes a simple yet powerful connection between structural disorder and spectral properties in glasses, opening a new avenue for advancing their studies.

\end{abstract}

\maketitle

\section{Introduction}
In recent years, machine learning (ML) methodologies have witnessed a swift and widespread uptake across the breadth of scientific inquiry. 
We have sought to leverage successful frameworks and methodologies from artificial intelligence to address complex problems across diverse disciplines.
For example, in condensed matter physics and related fields, neural network quantum states and wavefunctions, machine learning force fields, and regression models for structure-property relationships (SPR) have made influential contributions to many advances \cite{cheng2026artificial,kalita2025machine,tang2025deep,hermann2023ab,reiser2022graph}. 
These successful implementations generally rely on two core factors, namely access to large-scale, high-fidelity datasets, or the exceptional representational capacity of modern neural networks.
Yet alongside these advances, the community has growing concerns about the black-box nature of state-of-the-art models, which many argue hinders the discovery of new physical insights and principles \cite{hassija2024interpreting}. 
A further unresolved question is whether we have identified optimal architectures for different scientific tasks without overusing the advanced algorithms.

In this study, we focus on the SPR for glasses. 
Unraveling the connection between atomic structure and physical properties in such systems has long remained a significant challenge, owing to the inherent complexity of their non-crystalline nature, which renders the development of universal empirical descriptors inherently intractable \cite{anderson1995through}. While correlations may be proposed between specific properties and empirical metrics, e.g. medium-range order parameters \cite{angell1985spectroscopy,gaskell1996medium}, coordination numbers \cite{thorpe1983continuous,phillips1985constraint}, and packing fractions \cite{shen2009icosahedral,leocmach2012roles}, these relationships are often weak and could prove unreliable in practice \cite{biroli2008thermodynamic,berthier2011theoretical,tian2023disorder,cheng2024regulating}.
In recent years, as anticipated, more accurate quantitative SPRs have been established for glassy materials with the aid of ML architectures \cite{ravinder2021artificial,liu2025amorphous,han2025ai,jung2025roadmap}.
For example, unsupervised ML algorithms like topological data analysis and autoencoder have been utilized to uncover hidden structural signatures in glass \cite{sorensen2020revealing,yang2024structural}; supervised ML like deep convolutional neural networks and graph neural networks are capable of predicting a wide range of electrical, mechanical and thermodynamic properties of glass \cite{fan2021predicting,swanson2020deep,yang2024structural,alcobaca2020explainable}. 
However, despite their improved predictive performance, most existing ML-based SPR approaches for glasses still offer limited physical interpretability, providing little direct insight into how specific structural features quantitatively contribute to macroscopic properties.

Building on the recent progress, here we re-examine the SPR problem in glasses and demonstrate that an approximate linear theoretical relationship can be established between the radial distribution function (RDF) and the disorder-induced contributions to glassy physical properties. 
This linear theoretical framework facilitates the development of a simple linear regression ML model, which can be efficiently trained to predict SPRs with high accuracy.
We validate the effectiveness of this parsimonious model using amorphous monolayer carbon (AMC), periodic Lennard-Jones (LJ) systems, bulk amorphous SiC, and ternary amorphous CuAlZr alloys, highlighting its advantages over state-of-the-art neural network models. 
Furthermore, we conduct an interpretability analysis via regularization methods to verify its alignment with fundamental design principles.
Overall, compared to state-of-the-art ML models, our linear model not only boasts higher computational efficiency, requires less training data and is less prone to overfitting, but also exhibits superior interpretability.

\section{Theory and Model}
\subsection{Linear approximations from perturbation theory}

The fundamental distinction between glassy materials and crystalline solids lies in their structural disorder. 
Crystalline materials possess long-range ordering, which typically corresponds to the global or stable local minima on the potential energy landscape (denoted with an arbitrary function $U$).
In contrast, glasses are disordered and their atomic structures deviate from the local energy minima occupied by the corresponding crystalline materials.
This observation motivates a first-order perturbative description of glassy structures and properties, in which deviations from the crystalline reference are treated explicitly at the leading order. 
As a direct consequence, structural descriptors and derived physical quantities become expressible as first-order expansions with respect to the atomic displacements, giving rise to explicit linear relations between structure descriptors and properties. 
The detailed derivations are as follows.

We first consider the perturbation of atomic structures. For each atom $i$, its glassy coordinate $\mathbf{r}_i$ is perturbed from the corresponding crystalline reference $\mathbf{r}_i^{(0)}$ by $\Delta \mathbf{r}_i$, such that $\mathbf{r}_i = \mathbf{r}_i^{(0)} + \Delta \mathbf{r}_i$. 
To describe the structure of amorphous materials, the RDF $g(r)$ is often employed. 
$g(r)$ is the radial part of the pair distribution function $g(\mathbf{r})$:
\begin{equation}
    g(\mathbf{r})=\frac{1}{C}  \sum_{i\neq j} \delta(\mathbf{r}-\mathbf{r}_i + \mathbf{r}_j)
\end{equation}
where $C$ is a proper normalization factor. Within the first-order perturbation, it can be written as
\begin{equation}
g(\mathbf{r}) = g^{(0)}(\mathbf{r}) - \frac{1}{C} \sum_{i\neq j}\left[ \frac{\partial}{\partial \mathbf{r}_j} \delta(\mathbf{r} - \mathbf{r}_i^{(0)} + \mathbf{r}_j^{(0)}) (\Delta \mathbf{r}_i-\Delta \mathbf{r}_j) \right].
\label{eq:rdf}
\end{equation}
This gives a set of linear equations between $g(\mathbf{r})$ and $\{\Delta \mathbf{r}_i \}$,
hence a bi-direction linear mapping between $g(\mathbf{r})$ and spatial deviations exists.

For the property descriptor, we take the vibrational spectra as an example.
The aperiodicity of amorphous materials compresses the momentum space information to the origin of the Brillouin zone.
Hence the dynamic matrix that describes the phonon features is directly proportional to the Hessian matrix, 
and the phonon density of states (PDOS) is calculated as:
\begin{equation}
    D(\omega) = \sum_n \delta(\omega - \omega_n),
\end{equation}
where $\omega_n$ are phonon frequencies derived from the eigenvalues $\lambda_n=\omega_n^2$ of the Hessian matrix, given by $H_{ij}=\frac{\partial^2 U}{\partial\mathbf{r}_i\partial\mathbf{r}_j}$. 
Perturbations of $\Delta \lambda_n$ affects $\omega_n$, which further transfers to $D(\omega)$:
\begin{equation}
\begin{aligned}
    D(\omega) &= D^{(0)}(\omega)-\sum_n \frac{d}{d\omega}\delta(\omega-\omega_n^{(0)})\frac{1}{2\omega_n^{(0)}}\Delta\lambda_n.
\end{aligned}
\label{eq:dos}
\end{equation}

Meanwhile, one can also show that a linear mapping exists between $\lambda_n$ and structural perturbations $\Delta \mathbf{r}_k$. 
To derive this relation, we inspect the modulation of the structural disorder to the Hessian matrix is defined as:
\begin{equation}
\begin{aligned}
    H_{ij} &= H_{ij}^{(0)} + \sum_k \Phi_{ijk}\Delta \mathbf{r}_k,
    \end{aligned}
\end{equation}
where $\Phi_{ijk} = \frac{\partial{H_{ij}}}{\partial \mathbf{r}_k} $ denotes the third-order force constant that introduces anharmonic effects.
This relation demonstrates that structural disorder $\Delta \mathbf{r}_k$ perturbs the Hessian matrix elements, with the magnitude of perturbation directly linked to anharmonicity.

To proceed with the derivation, we further assume that the displacement vectors $\Delta \mathbf{r}_i$ for distinct atoms are independent Gaussian random variables, $\Delta \mathbf{r}_i \sim \mathcal{N}(0, \sigma_r^2)$, with the purpose of enabling a quantitative estimate of the resulting eigenvalue variations.
Under this assumption, the induced Hessian perturbation $\Delta H_{ij}=\sum_k \Phi_{ijk}^{(0)}\Delta \mathbf{r}_k$ can be treated as a random matrix with controlled variance.
We therefore invoke two standard results from random matrix theory~\cite{vershynin2018high,franklin2012matrix} to bound the associated eigenvalue deviations. \textit{Theorem 1 (Weyl's Inequality):} For a symmetric matrix $H_0$ and a perturbation $H$, the eigenvalue perturbation is bounded by $|\lambda_i(H_0 + H) - \lambda_i(H_0)| \leq \|H\|$, where $\|H\|=\text{max}_i|\lambda_i(H)|$ is the operator norm. \textit{Theorem 2 (Random Matrices with Gaussian Entries):} For an $n\times n$ random matrix $H$ with Gaussian entries of variance $\sigma^2$, for any $t>0$ with probability at least $1-2 e^{-t^2}$, the operator norm of $H$ satisfies the bound of $\|H\| \leq c \cdot \sigma (2\sqrt{n} + t)$, where $c$ is a constant. 
The first theorem guarantees eigenvalue stability under small Hessian matrix perturbations, and the second theorem confirms that the Gaussian Hessian perturbation $\Delta H_{ij}=\sum_k\Phi_{ijk}\Delta \mathbf{r}_k$ is strictly confined to a small magnitude proportional to $\sigma_r$ with a very high probability.
These results jointly establish that the eigenvalue shifts induced by structural disorder are uniformly controlled, thereby providing a rigorous justification for applying first-order perturbation theory to the phonon spectrum.

Within this perturbative framework, the eigenvalues of the Hessian matrix can be expressed as:
\begin{equation}
\begin{aligned}
     \lambda_n &= \lambda_n^{(0)} + \Delta \lambda_n,
\end{aligned}
\end{equation}
in which $\Delta \lambda_n=\langle n| \Delta H|n\rangle$ and $|n\rangle$ is the $n$-th eigenvector of the Hessian matrix $H^{(0)}$.
Given $\Delta H_{ij} \propto \Delta \mathbf{r}_k$, we reach the conclusion that the change of each eigenvalue $\lambda_n$ is approximately a linear transformation of the atomic coordinate perturbations $\Delta \mathbf{r}_k$.

Finally, we establish linear approximations all the way from RDF $g(r)$ to the PDOS $D(\omega)$. 
Notably, this linear approximation is independent of the specific form of the interatomic potential and applies broadly to glassy systems across different spatial dimensions.

\subsection{Linear SPR model}

\begin{figure*}
    \centering
    \includegraphics[width=0.95\linewidth]{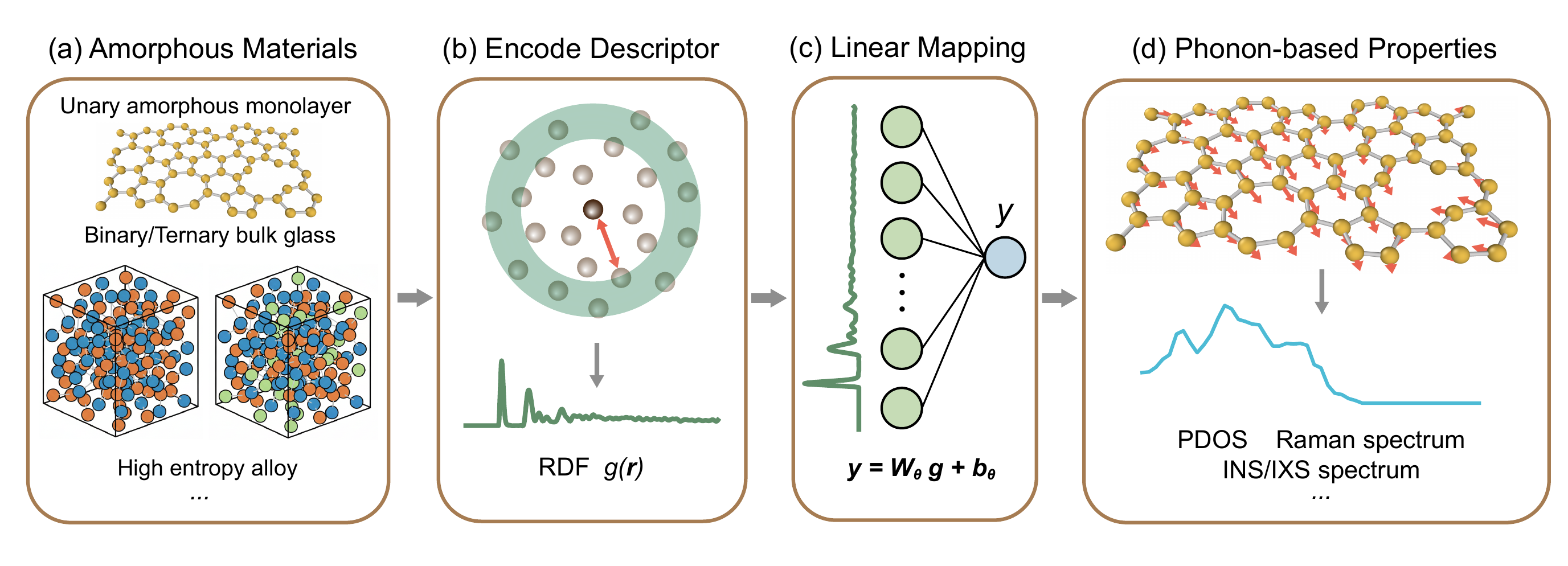}
    \caption{\textbf{Universal linear SPR for phonon-related properties of amorphous materials.} \textbf{(a)} Representative amorphous systems considered in this work, including unary amorphous monolayers, binary and ternary bulk glasses, and high-entropy alloys, spanning a wide range of structural disorder. \textbf{(b)} Atomic configurations are encoded using RDF $g(r)$, which captures two-body structural correlations in a symmetry-invariant and size-scalable manner. \textbf{(c)} A simple linear SPR model maps the vectorized $g(r)$ to target property features $y$ through learnable weight ($W_\theta$) and bias ($b_\theta$) parameters shared across all structures. \textbf{(d)} The resulting framework enables the prediction of multiple phonon-derived observables, such as PDOS, Raman spectrum, INS and IXS spectrum.}
    \label{fig1}
\end{figure*}
Leveraging the insights gleaned from our linear approximation framework, we formulate a simple linear SPR framework for glass, as is illustrated in Fig.\,\ref{fig1}.
As shown in Fig.\,\ref{fig1}(a), we consider a broad class of amorphous materials, ranging from unary amorphous monolayers to binary and ternary bulk glasses and high-entropy alloys, encompassing structural configurations that span from crystalline order to fully amorphous disorder. 

While atomic coordinates provide a complete representation of the structure, they reside in a $3N$-dimensional space and are therefore unsuitable for direct learning or interpretation. 
Motivated by the linear approximation framework developed in the previous section, we instead adopt a compact yet physically meaningful descriptor, the RDF $g(r)$ as the structural descriptor of glass.
$g(r)$ captures two-body correlations that are robust across amorphous systems, as is shown in Fig.~\ref{fig1}(b), while remaining scalable with system size and invariant under global translations, rotations, and permutations of identical atoms, making it an ideal input descriptor for ML-based property prediction.

Within this framework, we formulate a simple linear SPR model of the form (Fig.\,\ref{fig1}(c))
\begin{equation}
\label{eq:linear}
\mathbf{y} = \mathbf{W}_\theta \,\mathbf{g} + \mathbf{b}_\theta.
\end{equation}
Here $\mathbf{g}$ denotes the vectorized RDF, constructed by discretizing RDF $g(r)$ over radial bins and serving as the input structural feature; the output $\mathbf{y}$ represents the predicted property feature, which may correspond to a scalar quantity or a vectorized, symmetry-invariant representation of a physical observable.
The parameters $\mathbf{W}_{\theta}$ and $\mathbf{b}_{\theta}$ are learnable weight and bias terms, respectively, and are shared across all structures, encoding a universal linear mapping from structural descriptors to target properties.
Beyond the explicit linear relationship between RDF $g(r)$ and the PDOS $D(\omega)$ demonstrated earlier, our perturbative analysis establishes that small structural variations induce linear responses in the Hessian eigenvalues and lead to correspondingly smooth, perturbative changes in the associated eigenvectors, up to trivial phase ambiguities.
As a result, a broad class of phonon-derived observables that depend smoothly on the Hessian spectrum, such as thermal conductivity, Raman spectrum, inelastic neutron scattering (INS) and inelastic X-ray scattering (IXS) spectrum, are likewise expected to admit accurate linear mappings from $g(r)$ (Fig.\,\ref{fig1}(d), more detailed explanation is shown in the Discussion section). 
This linear SPR framework thus provides a unified and interpretable route for predicting multiple phonon-related properties of glass from a single, low-dimensional structural descriptor $g(r)$.

\section{Results}
To illustrate the effectiveness of the linear model, we perform a series of numerical experiments on diverse glassy systems, including AMC, periodic LJ systems, bulk amorphous SiC, and a ternary amorphous CuAlZr alloy.
These systems cover a wide range of chemical compositions and interatomic potentials, and collectively demonstrate a universal linear SPR, in which phonon properties are linearly mapped from the radial distribution function of amorphous materials.

\subsection{Amorphous monolayer carbon}

\begin{figure*}
    \centering
    \includegraphics[width=0.95\linewidth]{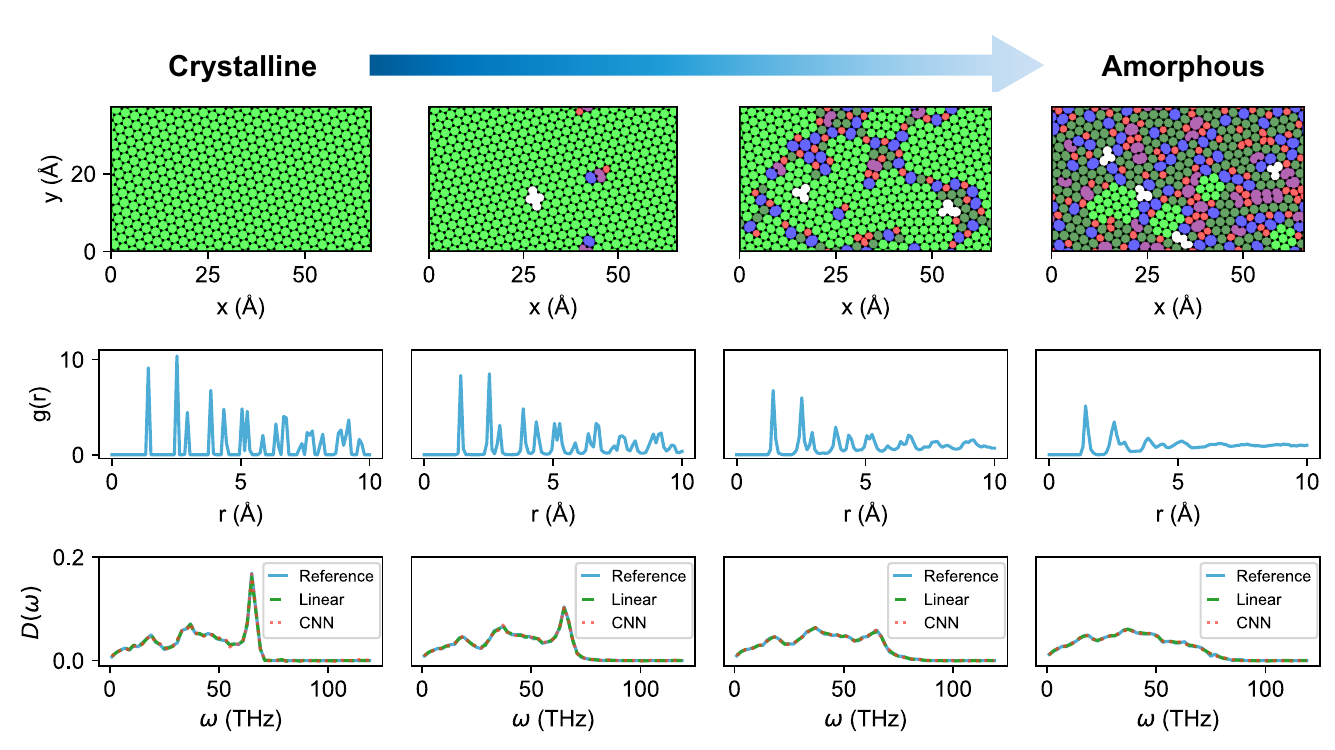}
    \caption{\textbf{Structures, RDF, and PDOS of AMC with varying disorder and PDOS prediction results.} Representative AMC structures with increasing degrees of disorder are shown from left to right, together with their corresponding RDF and PDOS. Columns span a structural continuum from crystalline-like to highly amorphous configurations. Atomic structures are colored according to ring statistics and local bonding environments: six-membered rings are shown in green (bright green denotes a six-membered ring that has at least one nearest-neighboring six-membered ring whose nearest-neighbors are all six-membered, while the remaining six-membered rings are colored dark green); five-, seven-, and eight-/nine-membered rings are shown in red, blue, and purple, respectively. Vacancies are rendered transparent. For each AMC structure, the PDOS predicted by a linear model and by CNN are compared against the reference PDOS obtained from calculations in the third row, illustrating the predictive performance of the two models across different levels of structural disorder.
}
    \label{fig:linear_SPR}
\end{figure*}

Two-dimensional amorphous materials provide an ideal platform for studying SPR because their atomic configurations can be directly visualized in experiment \cite{huang2013imaging,hong2020ultralow,joo2017realization,toh2020synthesis,tian2023disorder,bai2024nitrogen}. 
Among them, AMC is particularly useful because its degree of disorder can be tuned by synthesis conditions and because its electronic conductivity can vary by orders of magnitude with disorder \cite{toh2020synthesis,tian2023disorder}.
We therefore highlight AMC as a primary test bed for the linear SPR model in this work.
Previous work has developed SPRamNet, which can accurately predict electronic and thermal properties of AMC with $g(r)$ as the sole input feature \cite{cheng2025predicting}. 
While SPRamNet demonstrates the effectiveness of $g(r)$ in establishing the SPR of amorphous materials, it relies on training on a deep convolutional neural network (CNN), which introduces relatively high model complexity and offers limited interpretability. 
Here we demonstrate that this complexity can be directly stripped away, and a simple linear model suffices to map $g(r)$ to PDOS across a wide range of disorder in AMC.

We generate AMC structures with varied degree of disorder by molecular dynamics (MD) quenching simulation, using the empirical Tersoff potential \cite{tersoff1988empirical}. 
Two control parameters govern disorder in our dataset, namely the quench rate and the defect concentration. In general, faster quenching produces higher disorder; removing a fraction of atoms creates vacancy-like defects and voids. 
Further details of structure generation are given in the Supplementary Information I, following Ref.\,\cite{cheng2025predicting}. 
Several typical structures from our dataset are shown in the top row of Fig.\,\ref{fig:linear_SPR}. 
The dataset spans a broad disorder range, including fully ordered crystalline graphene, graphene distorted by vacancies and Stone-Wales (SW) defects, and highly disordered AMC.

For each structure we compute its RDF and PDOS(Details of calculating RDF and PDOS are given in Supplementary Information II). 
RDF and PDOS for representative structures are shown in Fig.~\ref{fig:linear_SPR}, illustrating how disorder systematically modifies both structural and vibrational signatures. 
In the RDF, disorder leads to a clear broadening of bond-length distributions compared to the crystalline state. Peaks at short distances ($r< 4\,\rm{\mathring{A}}$) remain relatively sharp, indicating preserved short-range order, whereas peaks at larger distances are rapidly smoothed as disorder increases, consistent with the loss of medium- and long-range order in AMC. 
Correspondingly, disorder affects different frequency ranges of the PDOS of pristine graphene to significantly different extents. 
The crystalline PDOS exhibits three pronounced features near 19\,THz, 38\,THz, and 65\,THz. 
As disorder increases, the high-frequency feature around 65\,THz, dominated by bond-stretching optical vibrations, is most strongly affected, undergoing substantial broadening and progressive suppression until it becomes nearly indistinguishable; on the other hand, the lower-frequency features near 19\,THz and 38\,THz, which involve more collective acoustic or mixed acoustic-optical motion, exhibit comparatively modest changes.

We test the linear SPR defined in Eq.~\ref{eq:linear} for predicting PDOS from RDF. 
The simple linear model is evaluated against the CNN previously developed in SPRamNET \cite{cheng2025predicting}, and the results are shown in the bottom row of Fig.\,\ref{fig:linear_SPR}. 
Surprisingly, both models yield PDOS predictions that are visually indistinguishable from each other and in excellent agreement with the reference PDOS, with all three curves overlapping across the entire frequency range.
More quantitatively, the linear model achieves a training loss of $1.7\times 10^{-6}$, comparable to $1.4\times 10^{-6}$ for the CNN(See Supplementary Information III). 
This demonstrates that a linear mapping would suffice to capture the SPR between RDF and vibrational spectra in AMC.

\begin{figure}
    \centering
    \includegraphics[width=1.0\linewidth]{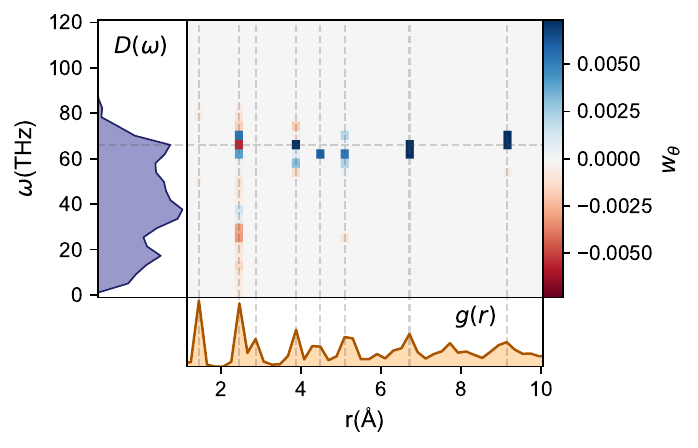}
    \caption{\textbf{Visualization of weights for the linear SPR mapping of amorphous monolayer carbon.} Learned linear weights are shown in a heat map for a representative AMC structure, using L1 and L2 regularized weights $\alpha=10^{-5},\ \beta=0$, respectively. The colored regions indicate non-zero weights, and the dashed lines indicate the correspondence between the peaks of RDF and PDOS. The reference RDF and PDOS showed above are corresponding to a moderate disorder structure.}
    \label{fig:regularzation}
\end{figure}

With minimal loss of accuracy or expressibility, the linear model additionally offers the advantage of interpretability. 
A straightforward diagnostic is to examine the learned weight matrix $\mathbf{W_\theta}$, which reveals which RDF regions most strongly influence each PDOS frequency. 
To obtain sparser and more interpretable weights, we include L1 and L2 regularization in the loss function (between predicted $\hat{y}_\theta$ and labelled data $\hat{y}$) and minimize 
\begin{equation}
\mathcal{L}=\|\hat{y}-\hat{y}_\theta\|^2+\alpha \|\theta\|_1+\beta \|\theta\|_2^2
\end{equation}
where $\alpha$ and $\beta$ are the L1 and L2 regularization parameters, respectively, and $\theta$ represents all learnable parameters. 
A parameter grid search indicates that small regularization ($\alpha \lesssim 10^{-5},\ \beta \lesssim 10^{-4}$) has minor impact on the prediction loss (see Supplementary Information IV), but will substantially enhance the interpretability of the learned weights.

For example, Fig.~\ref{fig:regularzation} shows the result for $\alpha=10^{-5},\ \beta=0$ for a representative SPR mapping, where L1 regularization yields a sparse set of non-linear weights.
These weights align closely with prominent features in both the RDF and PDOS, as indicated by the dashed lines.
In the low-frequency regime, the PDOS is primarily governed by short-range structural information, with the dominant contributions arising from RDF features at $r = 2.5\rm{\mathring{A}}$. 
By contrast, high-frequency regions, such as the pronounced PDOS feature near $65\mathrm{THz}$, are predominantly associated with weights at larger interatomic distances in the RDF. 
This indicates an increased sensitivity of high-frequency vibrational modes to medium- and longer-range structural correlations. 
Besides weights, we also show the change of bias after imposing regularization in Supplementary Information IV. 
In contrast to the chaotic lines produced without regularization, the bias with regularization closely matches the PDOS in both shape and magnitude at low frequencies. 
This agreement is consistent with the fact that disorder induces relatively minor changes to the PDOS in the low frequency spectral region of AMC. 
These observations demonstrate that the learned linear mapping encodes a physically meaningful, scale-dependent relationship, in which vibrational features at different frequencies are selectively influenced by structural correlations at distinct length scales.

\subsection{Extension to other glassy systems}

\begin{figure*}
    \centering
    \includegraphics[width=0.95\linewidth]{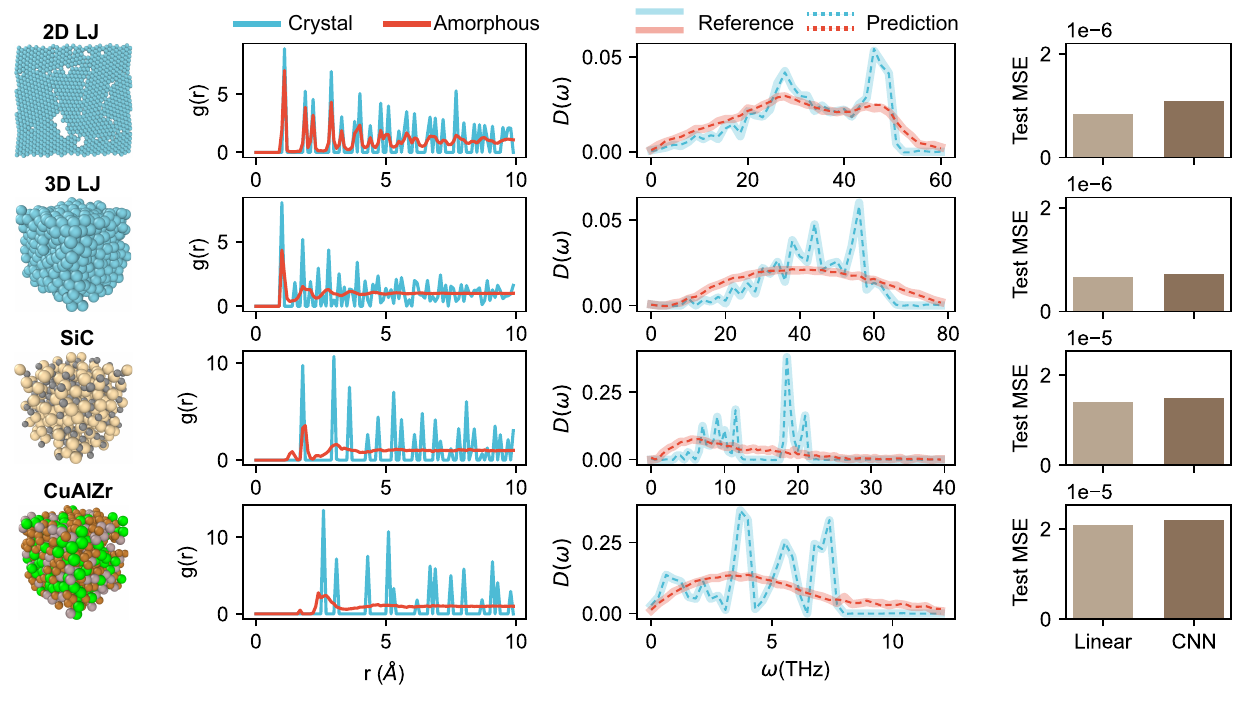}
    \caption{\textbf{Linear structure–property relation across disordered systems of increasing complexity.} Representative atomic configurations (first column), RDF ($g(r)$) (second column), and corresponding PDOS (third column) are shown for four disordered systems: 2D LJ, 3D LJ, binary covalent SiC, and ternary metallic CuAlZr. 
    Red curves correspond to the highly disordered structures shown in the first column, and the blue curves represent the RDF and PDOS of the crystalline phases.
    Solid curves denote reference data, while dashed curves indicate predictions of the linear model. The rightmost column compares test MSE of the linear model and CNN. }
    \label{fig:more_example}
\end{figure*}

In the previous section we have established that the linear SPR is valid and physically interpretable for the vibrational spectra of AMC. 
While AMC serves as an ideal test bed for SPR mapping, several open questions remain. 
First, AMC is a two-dimensional material, so it is unclear whether the linear SPR also applies in three-dimensional systems. 
Second, our AMC simulations use a three-body empirical potential, and it is important to verify the SPR with both simpler two-body potentials and with more complicated, many-body machine-learned interatomic potentials (MLIPs). 
Third, AMC is a single-component material (carbon), so multi-component amorphous systems require additional validation. 

Motivated by these possible extension beyond AMC, in this section we test the linear SPR across a variety of additional systems with different interatomic potentials under the periodic boundary condition, including the two-dimensional Lennard-Jones solid (2DLJ), its three-dimensional analogue (3DLJ, e.g., bulk Ar), amorphous SiC modeled with a pre-trained MACE MLIP \cite{batatia2022mace,batatia2025foundation}, and a ternary amorphous alloy CuAlZr modeled with embedded-atom method (EAM) many-body potential\cite{cheng2009atomic}. 

The four systems studied here were purposefully selected to span a wide range of material characteristics. 
The 2DLJ/3DLJ systems use the simplest pairwise interactions on calculating vibrational properties, providing a minimal reference for how packing and coordination affect the linear SPR. 
Besides, the LJ potential has been widely used as a model in the dynamical study and machine-learning analysis of SPR in amorphous materials \cite{kob1997dynamical,cubuk2015identifying,fan2021predicting}. 
We therefore choose the LJ system as a representative platform to assess the robustness and generality of the linear SPR approach. 
In contrast, the covalently bonded SiC system is modeled with a MACE-MP0 MLIP \cite{batatia2025foundation}, which predicts atomic energies and forces through a message-passage graph neural network.
The message-passing scheme explicitly encodes multi-atom correlations into its neural network layers, making MACE a many-body potential well suited for capturing the directional bonding and pronounced short-range order characteristic of covalent networks.
Finally, the ternary metallic alloy CuAlZr, modeled by the EAM potential\cite{cheng2009atomic}, is the basis of many important bulk metallic glasses with interesting properties, and introduces compositional complexity and chemical disorder characteristic of multi-component glasses. 
Therefore, these choices above cover a wide range of interaction physics (pairwise vs. many-body), bonding character (metallic vs. covalent), dimensionality, and chemical complexity, enabling a stringent test of whether linear SPR can be reliably generalized as a universal relation beyond a single case study of AMC.

All four systems' structures are generated with a consistent protocol: starting from the most stable crystalline structures, we apply randomized atomic displacements, remove a controlled fraction of atoms to introduce vacancies, and perform energy minimization. 
Such a protocol is broadly applicable to the systems studied here without the need for careful parameter tuning. 
By varying the displacement amplitude and the vacancy fraction we obtain structures that continuously span from mildly distorted crystals to highly amorphous configurations. 
Compared to quench MD simulation for AMC, this procedure also avoids long, computationally expensive MD runs. 
It's an important practical advantage when using costly potentials such as MACE, and therefore substantially improves sampling efficiency. 
Further details on structure generation are given in the Supplementary Information I.

Representative disordered structures generated by this protocol are shown in the first column of Fig.~\ref{fig:more_example}. The corresponding RDF and PDOS, together with the predictions from the linear model, are displayed in the second and third rows of Fig.~\ref{fig:more_example}, respectively. 
These rows present the results for highly disordered structures and the crystalline reference. 
Structures with moderate disorder and their corresponding properties are provided in Supplementary Information II.

For both the two-dimensional and three-dimensional periodic LJ systems, we observe the same qualitative trends as in AMC: increasing disorder leads to a progressive smoothing of RDF peaks and a broadening or suppression of high-frequency features in the PDOS.
Similar high-frequency broadening behavior is also observed in SiC and CuAlZr. 
However, in contrast to the single-element LJ systems, these multi-component materials exhibit additional low-frequency vibrational features that warrant special attention. 
Notably, in amorphous SiC under strongly disordered conditions a distinct peak appears smaller than 10 THz; previous work attributes such low-frequency features to chemical disorder, like the formation of Si-Si bonds that are absent in crystalline SiC. \cite{ivashchenko2007simulations} This type of PDOS modification induced by chemical disorder is characteristic of non-unitary amorphous materials and is not observed in elemental systems.

Despite the increased complexity of PDOS evolution introduced by chemical disorder, the linear SPR continues to perform robustly across all systems considered in this study. 
As is shown in Fig.\,\ref{fig:more_example}, the linear model accurately reproduces the reference PDOS for all four systems, with discrepancies that are barely discernible.
The fourth column of Figure~\ref{fig:more_example} compares the prediction loss of the linear model and the CNN for each system. 
As expected, the average mean squared error (MSE) overall increases with the chemical complexity, yet remains well controlled, reaching only $\sim 2\times 10^{-5}$ for the most complex ternary alloy CuAlZr. 
Moreover, the linear model achieves performance comparable to, and in some cases even better than, that of the CNN over a wide range of disorder parameters. 
A plausible explanation is that, although the CNN is more expressive, it requires substantially larger datasets to robustly learn high-frequency and system-specific nonlinearities, making it more susceptible to overfitting when the training data has a limited volume. 
This behavior is also evident in the training histories for AMC (see Supplementary Information I and III).
In contrast, the linear model has fewer degrees of freedom, requires substantially less training data, and is stabilized from regularization, leading to improved generalization performance.
All these results above suggest that, for vibrational properties in glass, linear SPR already capture the dominant physics across materials of widely varying complexity.

\section{Discussion}
In this work, we demonstrate that a simple, regularized linear model provides an accurate and interpretable SPR that maps RDF to PDOS across a broad class of disordered systems.
Surprisingly, this linear SPR consistently captures the dominant trends with accuracy comparable to, or even exceeding, that of more expressive CNN models, when limited data is available for training. 
Together with the visualized, interpretable linear weights, these results suggest that for vibrational properties, the essential SPR is predominantly linear and governed by disorder-driven smoothing of structural correlations.

Beyond the PDOS, the linear SPR framework is likely applicable to a broader class of vibrational observables as well. 
As shown in Sec.~II, phonon-related observables that depend smoothly on the Hessian spectrum are generally expected to follow an approximately linear structural–property relationship (SPR) with respect to the radial distribution function (RDF). 
Such phonon-related observables are by no means limited to the PDOS; rather, they encompass many other properties, particularly those accessible to experimental measurements, thereby providing substantial opportunities for experimental validation.

The first category includes observables that depend primarily on the atomic structure. 
One example is the thermal conductivity of amorphous insulating materials, which is dominated by phonon transport and can be measured experimentally and calculated theoretically within the Allen-Feldman formalism~\cite{allen1993thermal}. 
This approach requires the vibrational density of states $D(\omega)$, the mode heat capacity $C_\nu$, and the mode diffusivity $D_\nu$, all of which are determined by the vibrational frequencies and modes. 
Other widely used experimental probes in this category include INS and IXS. 
These techniques have been extensively employed to characterize the static structures of amorphous materials~\cite{tulk2002structural,laaziri1999high,gardner1996structural} as well as phonon properties~\cite{kamitakahara1984measurement}. 
From INS and IXS measurements, one can extract the dynamic structure factor $S(q,\omega)$, whose theoretical evaluation likewise relies on the phonon frequencies $\{\omega_\nu\}$ and modes $\{\mathbf{e}_\nu\}$ through the $q$-resolved vibrational density of states.

Beyond observables predominantly determined by atomic structure, there also exists a class of properties governed jointly by atomic and electronic degrees of freedom, such as infrared (IR) absorption spectra (dipole activity) and Raman spectra (polarizability fluctuations). 
These spectroscopies are particularly important for amorphous materials: unlike crystalline systems, where only a limited number of symmetry-allowed modes are optically active, the absence of long-range periodicity and the breaking of translational and point-group symmetries in amorphous states render nearly all vibrational modes optically excitable. 
As a result, IR and Raman spectroscopies serve as widely used experimental probes of vibrational excitations in amorphous materials~\cite{shuker1970raman,elliot582446368physics}.
The IR intensity depends on the phonon frequencies and the Born effective charges $\partial \boldsymbol{\mu}/\partial Q_\nu$, whereas Raman scattering involves the Raman tensor $\partial \boldsymbol{\alpha}/\partial Q_\nu$. 
Although both quantities depend explicitly on the electronic structure, previous studies have shown that in the nonresonant and weak-disorder limit, variations in the Raman tensor are subdominant, and the spectral lineshape is primarily governed by the vibrational eigenfrequencies~\cite{hashemi2019efficient}. 
Under these conditions, a linear SPR is therefore still expected to hold approximately. 
In contrast, IR absorption intensities generally cannot be expressed as a linear functional of the PDF alone, owing to their stronger sensitivity to local electronic polarization effects.

The above discussions are summarized in Table~\ref{tab:exp_probes_prx}. 
From an experimental perspective, the existence of a robust and approximately linear SPR enables a quantitative mutual convertibility between real-space structural measurements and vibrational spectroscopies. 
In principle, vibrational spectra can be used to infer effective structural correlations encoded in the PDF, while diffraction-derived PDFs may be employed to predict vibrational responses. 
Such cross-modal inference, enabled by the linear nature of the SPR, offers enhanced experimental flexibility, reduced characterization costs, and valuable internal consistency checks across complementary experimental probes.

\begin{table*}[t]
\caption{Experimental probes connecting real-space structure and vibrational properties in amorphous materials. For each technique we list the primary measured observable, the quantities required for theoretical evaluation, and whether a linear mapping is expected to be applicable.}
\label{tab:exp_probes_prx}
\begin{ruledtabular}
\begin{tabular}{p{2.1cm} p{4.2cm} p{5.8cm} p{2.6cm}}
Technique
& Measured observable
& Required quantities for simulation
& Applicability \\ \hline
Inelastic neutron scattering (INS)
& Dynamic structure factor $S(Q,\omega)$; $Q$-averaged spectra yield neutron-weighted $D(\omega)$
& Phonon frequencies and modes $\{\omega_\nu,\mathbf{e}_\nu\}$
& Yes \\

Inelastic X-ray scattering (IXS)
& Dynamic structure factor $S(Q,\omega)$ from coherent density fluctuations
& Phonon frequencies and modes $\{\omega_\nu,\mathbf{e}_\nu\}$
& Yes \\

Raman spectroscopy
& Frequency-resolved scattering intensity $I(\omega)$ from polarizability fluctuations
& Phonon frequencies and modes $\{\omega_\nu,\mathbf{e}_\nu\}$ and Raman tensor $\partial\boldsymbol{\alpha/}\partial Q_\nu$
& Yes$^{\ast}$ \\

Infrared spectroscopy (IR)
& Absorption intensity from dipole-active vibrational modes
& Phonon frequencies and modes $\{\omega_\nu,\mathbf{e}_\nu\}$ and Born effective charges $\partial\boldsymbol{\mu/}\partial Q_\nu$
& No \\

Thermal transport
& Thermal conductivity $\kappa(T)$
& Vibrational density of states $D(\omega)$, mode heat capacity $C_\nu$, mode diffusivity $D_\nu$
& Yes$^{\ast\ast}$ \\

\end{tabular}
\end{ruledtabular}
\begin{flushleft}
$^{\ast}$Approximate linearity applies in the nonresonant, weak-disorder limit.\\
$^{\ast\ast}$Phonon contribution only.
\end{flushleft}
\end{table*}

Beyond the diversity of output observables, the linear SPR framework is also not restricted to a specific choice of structural descriptor as input. 
As discussed in Supplementary Information~V, any descriptor whose features broaden or smoothen monotonically with increasing structural disorder—such as angular distribution functions or topological descriptors—may serve as a suitable linear input. 
Recent work employing persistent homology as a disorder-aware structural descriptor has successfully predicted the thermal conductivity of amorphous carbon using ridge regression \cite{yamazaki2025topological}, providing concrete evidence that linear SPRs can emerge for a broad class of vibrational properties when appropriate structural representations are adopted.

Despite these advances, two challenges still exist.
Firstly, other properties beyond vibrational properties remain to be explored. For example, the electric conductivity and electron density of states in AMC have been shown to depend sensitively on additional factors beyond the smoothening of pair correlations, and the SPR cannot be reliably reconstructed using a shallow neural network, let alone a linear mapping \cite{cheng2025predicting}. 
Therefore, constructing an explainable SPR for electronic transport likely requires alternative descriptors or explicitly nonlinear mappings and merits dedicated investigation.
Additionally, while quenching, randomized distortions, and atom removal sample a diverse ensemble of amorphous configurations, they do not exhaustively capture all experimentally relevant microstructure or defect motifs, and may not fully cover the vast configurational space of disordered structures. 
Incorporating more realistic defect formation pathways or experimental structure ensembles would be an important direction for future work.

Looking ahead, our linear SPR framework suggests a promising direction for modeling disordered materials, in which physical insights and predictive power are achieved not through increasing model complexity and data volume in deep learning, but through identifying the appropriate disorder-aware structural descriptors and the most interpretable mapping. 
The linear SPR framework provides a foundation for systematically extending interpretable SPR to a broader class of experimental observables and amorphous material systems.
More broadly, our results point to a unifying perspective in which much of the complex physics of disordered materials may likely be captured by simple, physically grounded linear relations, offering a scalable route toward data-efficient prediction and materials discovery in glassy systems, and beyond.

\section{Acknowledgments}
This work was supported by the National Science Foundation of China under Grant No. 52541026. 
We are grateful for computational resources provided by the High Performance Computing Platform of Peking University.

\bibliography{refs.bib} 
\end{document}